\documentclass[aps,prd,preprint,amsmath,amssymb,nofootinbib,eqsecnum,showpacs]{revtex4}
\usepackage{epsfig}
\newcommand{\be}{\begin{equation}}
\newcommand{\ee}{\end{equation}}
\newcommand{\bea}{\begin{eqnarray}}
\newcommand{\eea}{\end{eqnarray}}
\begin{document}
\title{Calculating Casimir Energies in Renormalizable Quantum Field Theory}
\author{Kimball A. Milton}
\email{milton@nhn.ou.edu}
\homepage{www.nhn.ou/
\affiliation{Department of Physics and Astronomy, University of Oklahoma,
Norman, OK 73019-0430}

\date{\today}
\preprint{OKHEP-02-05}

\begin{abstract}
Quantum vacuum energy has been known to have observable consequences
since 1948 when Casimir calculated the force of attraction
 between parallel uncharged
plates, a phenomenon confirmed experimentally with ever increasing precision.
Casimir himself suggested that a similar attractive self-stress existed for
a conducting spherical shell, but Boyer obtained a repulsive stress.
Other geometries and higher dimensions have been considered over the years.
Local effects, and divergences associated with surfaces and edges were
studied by several authors. Quite recently, Graham et al.~have re-examined
such calculations, using conventional techniques of perturbative quantum
field theory to remove divergences, and have suggested that previous
self-stress results may be suspect.  Here we show that the examples
considered in their work are misleading; in particular, it is well-known
that in two space dimensions a circular boundary has a divergence in the Casimir 
energy for massless fields, while for general spatial
 dimension $D$ not equal to an 
even integer the corresponding Casimir energy arising from massless
fields interior and exterior to a hyperspherical shell is finite.
It has also long been recognized that the Casimir energy for massive
fields is divergent for curved boundaries. These conclusions are reinforced by
a calculation of the relevant leading Feynman diagram in $D$  and in
three dimensions.
There is therefore no doubt of the validity of the conventional finite
Casimir calculations.
\end{abstract}

\maketitle

\section{Introduction}
\label{Sec1}
The Casimir effect remains one of the least intuitive consequences of
quantum field theory, and stands rather outside the usual development
of renormalization theory.  This is because it is inherently nonperturbative,
in that macroscopic boundary conditions or backgrounds cannot be
easily mimicked by perturbative interactions.  Its origins go back to
the very beginnings of quantum mechanics, because it can be thought of
as the change in the zero-point energy when the background is introduced.

After examining the van der Waals interaction between two
molecules and between a molecule and a conducting plate \cite{casimirandpolder},
Casimir was challenged by Bohr \cite{casimir50,milonni}  to
interpret this interaction in terms of zero-point energy \cite{casimircolloq}, 
and then to
recognize that the zero-point fluctuations of the electromagnetic field
implied a force between two such plates \cite{casimir}.  
The attractive nature of this
force was obviously consistent with the action-at-a-distance interpretation
of it as due to the attraction between fluctuating dipoles making up the
material of the plates.  But intuition flew out the window when Boyer
discovered that the energy, and hence the self-stress, on a perfectly
conducting spherical shell of zero thickness was positive or repulsive
\cite{boyersphere}.
Later, it was found that a cylinder was intermediate, giving rise to a
small but attractive force \cite{deraadcyl}.

Dimensional dependence was also dramatic.  Sen examined a circular boundary
in two dimensions and found that the energy was infinite\footnote{Unfortunately,
 the author had apparently forgotten
this divergence in Ref.~\cite{mcsce2}, wherein an attempt was
made to extract a finite Casimir energy for a circular
boundary.  The error was pointed out in Ref.~\cite{lesed2,nestcyl}.} 
\cite{sen,sen2}.  This was
later found to be part of a pattern: For a hyperspherical shell 
in $D$ spatial dimensions, the Casimir energy of a massless scalar field
was finite except when $D$ was a positive even integer, where the
energy or stress exhibits a simple pole \cite{benmil}.  
(For $D\le0$, branch points
occur at the integers.)  An intuitive explanation of this,
and of the corresponding sign changes, is still lacking.

Deutsch and Candelas were the first to examine the local effects of
fluctuating fields \cite{deutsch}, for other than the geometry of
parallel planes, which was considered by Brown and Maclay a decade
earlier \cite{brown}.  
Typically, surface divergences occur near boundaries,
although for flat boundaries with conformally-coupled fields, those divergences
disappear.  The reason the global Casimir energy of a (hyper)sphere is
finite is that there is a perfect cancellation between the interior
and exterior divergences. 
 This perfect cancellation is spoiled if the
shell has finite thickness, or if the speed of light is different on the
two sides of the boundary \cite{miltonballs}.  
Giving the fluctuating field a mass also
yields an unremovable divergence \cite{blau,borelkirles,elborkir,scancyl}
 except for the case of  plane boundaries.

Recently, Graham et al.~\cite{graham0,graham,graham2} 
have questioned these findings.  They have
developed an approach in which idealized boundary conditions are replaced
with interactions with an external (nondynamical) field.  Potentially
divergent terms are subtracted and replaced by perturbatively calculable
Feynman diagrams. After renormalization of these diagrams, the limiting
case when the external field becomes a delta function is taken.  In this way 
the results for the $D=1$ force
 are reproduced; but the authors find those finite
results rather unsatisfactory, so they discuss how their limiting procedure
gives rise to a different energy, corresponding, 
however, to the conventional force.
Then they turn to a circular boundary in two space dimensions
 and find that the Casimir energy is divergent; the implication is
that this is a general feature, so that all calculations of Casimir self-stress
are called into question.

However, as we remarked above, $D=2$ is a singular point.  What is called
for is a calculation for general $D$.  That is the purpose of this paper.
For simplicity, our attention will be restricted to scalar fields.
We will first, in Sec.~\ref{Sec2}, re-examine the $D=1$ calculation,
and show that the force is completely finite, while the energy density,
or more generally, the stress tensor, has a constant divergent part which
would be present if the boundaries were not present, and is therefore quite
without observable consequence.  For general $D$, unphysical surface
divergences appear in the stress tensor (unphysical because they do not
contribute to the stress on the sphere), which, for zero mass, vanish
if the conformal stress tensor is used.
  Then, in Sec.~\ref{Sec3}, we re-examine
the self-stress on a sphere in three dimensions, using time-splitting
to regulate the divergences.  The result is, once again, unambiguously
finite.  The critical calculation is given in Sec.~\ref{Sec4}, where we
review and simplify the diagrammatic subtraction method, and explicitly
compute the graph in which two external fields are inserted, in $D$ spatial
dimensions.  As expected, the result is divergent at $D=2, 4, 6, \dots$, 
but is otherwise finite for $D>3/2$.  We verify the finiteness for three
dimensions by directly calculate the oversubtracted graph for $D=3$,
where we see that the divergences in $E$ are independent of the radius
of the sphere.
Concluding remarks are offered in Sec.~\ref{Sec5}.
Some discussion of how dimensional continuation extracts the part of Feynman
graphs which contains the dependence on physically relevant parameters is
given in the Appendix.

\section{Casimir Effect for Dirichlet Plates}
\label{Sec2}
\subsection{Massless Scalar in 1+1 Dimensions}
\label{Sec2.1}
We begin by reconsidering the Casimir effect for a massive scalar field
which vanishes on two parallel plates (Dirichlet boundary conditions).
Although these considerations are familiar, and are given in some
detail in Ref.~\cite{miltonbook}, we will concentrate on the local
effect in 1+1 dimensions in order to make the divergence structure
manifest and make contact with the work of Graham et al.~\cite{graham2}.

For a massless scalar field $\phi$, the stress tensor is
\be
T^{\mu\nu}=\partial^\mu\phi\partial^\nu\phi-\frac12g^{\mu\nu}\partial_\lambda
\phi\partial^\lambda\phi.\label{stresstensor}
\ee
It will be noted that for one spatial dimension, this canonical tensor
coincides with the conformal one,
\be
T^\mu{}_\mu=0.
\ee
The scalar field satisfies the free equation
\be
-\partial^2\phi=0,
\ee
but is subject to the Dirichlet boundary conditions on the plates
at $x=0$ and $x=a$:
\be
\phi(x=0)=\phi(x=a)=0.
\ee
The corresponding Green's function satisfies
\be
-\partial^2G(x,t;x',t')=\delta(x-x')\delta(t-t'),
\ee
and
\be
G(0,t;x',t')=G(a,t;x',t')=0.
\ee
Since the Green's function is translationally invariant in time, it
is natural to introduce a corresponding Fourier transform,
\be
G(x,x';t-t')=\int_{-\infty}^\infty\frac{d\omega}{2\pi}e^{-i\omega(t-t')}
g(x,x';\omega);
\ee
the reduced Green's function satisfies the ordinary differential equation
\be
-\left(\omega^2+\frac{d^2}{dx^2}\right)g(x,x';\omega)=\delta(x-x').
\ee
We only need the solutions of this equation in two regions:
\begin{subequations}
\bea
0\le x,x'\le a:\quad g(x,x';\omega)&=& -\frac{\sin \omega x_<\sin\omega(x_>-a)}
{\omega\sin\omega a},\label{ingf}\\
a\le x,x':\quad g(x,x';\omega)&=&
\frac1\omega\sin\omega(x_<-a)e^{i|\omega|(x_>-a)}.\label{outgf}
\eea
\end{subequations}
Here $x_>$ ($x_<$) is the greater (lesser) of $x$ and $x'$.
These are to be compared to the free Green's function, when no plates
are present:
\be
g_0(x,x';\omega)=\frac{i}{2|\omega|}e^{i|\omega||x-x'|}.
\label{freegreen}
\ee

When we recognize that the Green's function is the time-ordered product
of the fields,
\be
\langle \phi(x,t)\phi(x',t')\rangle=\frac1iG(x,t;x',t')
\ee
we see that the vacuum expectation value of the stress tensor may be
obtained by applying a differential operator to the Green's function,
and then taking the spacetime points to be coincident.
For the $00$ component, that is, the energy, that differential
operator is
\be
\partial_0\partial'_0+\frac 12\partial^\lambda\partial'_\lambda
=\frac12\partial_0\partial_0'+\frac12\partial_x\partial'_x,
\ee
and so we obtain between the plates
\bea
\langle T^{00}\rangle&=&\int\frac{d\omega}{2\pi}\frac1{2i}(\omega^2+\partial_x
\partial'_x)g(x,x';\omega)\bigg|_{x=x'}\nonumber\\
&=&\int\frac{d\omega}{2\pi}\frac{\omega^2}2\frac{i}{\omega\sin\omega a}
[\sin\omega x\sin \omega(x-a)+\cos\omega x\cos\omega(x-a)]\nonumber\\
&=&\int\frac{d\omega}{2\pi}\frac{i\omega}2\cot\omega a\nonumber\\
&\to&-\frac1{4\pi}\int_{-\infty}^\infty d\zeta\,\zeta\coth\zeta a.
\label{unsubt00}
\eea
Here, in the last step we have made the complex frequency rotation,
\be
\omega\to i\zeta.
\ee
We notice that this last integral in Eq.~(\ref{unsubt00})
 does not exist.  This is because for
large $\zeta$ the hyperbolic cotangent approaches unity.  If we subtract
off this limiting value we obtain a finite result:
\bea
\langle T^{00}\rangle&\to&-\frac1{2\pi}\int_0^\infty d\zeta\,\zeta(\coth\zeta a
-1)\nonumber\\
&=&-\frac1\pi\int_0^\infty \zeta\,d\zeta \frac1{e^{2\zeta a}-1}\nonumber\\
&=&-\frac\pi{24 a^2}.
\eea
The energy is obtained from this by multiplying by the distance between
the plates:
\be
E=-\frac\pi{24 a},
\ee
which is the well-known L\"uscher potential \cite{luscher}.

In the same way we can calculate the vacuum expectation value of the
$xx$ component of the stress.  The relevant differential operator
\be
\partial_x\partial'_{x}-\frac12\partial_\lambda\partial^{\prime\lambda}
\to\frac12(\omega^2+\partial_x\partial'_{x})
\ee
is unchanged, so we obtain the same result as for $\langle T^{00}\rangle$.
The off-diagonal terms $\langle T^{0x}\rangle$ result from the application
of the symmetric differential operator 
\be
\frac12(\partial^0\partial^{\prime x}+\partial^x\partial^{\prime0}),
\ee
so are necessarily zero.  Keeping the divergent term we subtracted off,
the result for the stress tensor between the plates, $0\le x,x'\le a$,
is
\be
\langle T^{\mu\nu}\rangle=\left[u_{\rm vac}-\frac\pi{24 a^2}\right]
\left(\begin{array}{cc}
1&0\\
0&1\end{array}\right),
\ee
where the divergent terms is
\be
u_{\rm vac}=-\frac1{2\pi}\int_0^\infty d\zeta\,\zeta.
\ee
Note that $\langle T^{\mu\nu}\rangle$ is traceless,
\be
\langle T^\mu{}_\mu\rangle =0,
\ee
as required by conformal symmetry.

If we follow the same operations to find the stress tensor outside
the plates from Eq.~(\ref{outgf}) we obtain
\bea
\langle T^{00}\rangle=\langle T_{xx}\rangle&=&\frac1{2i}\int\frac{d\omega}
{2\pi}\frac1\omega\left[i\omega|\omega|\cos\omega(x-a)e^{i|\omega|(x-a)}
+\omega^2\sin\omega(x-a)e^{i|\omega|(x-a)}\right]\nonumber\\
&=&\frac1{4\pi}\int_{-\infty}^\infty d\omega\,|\omega|=-\frac1{2\pi}
\int_0^\infty d\zeta\,\zeta=u_{\rm vac}. 
\eea
That is, in the two regions outside the plates, $x<0$ or $x>a$,
\be
\langle T^{\mu\nu}\rangle =u_{\rm vac}
\left(\begin{array}{cc}
1&0\\
0&1\end{array}\right).
\ee
This is exactly the stress tensor that would be found everywhere
if the free Green's function $g_0$ in Eq.~(\ref{freegreen}) were used.
This means that the force on one of the plates is completely finite
and unambiguous, because it is given by the discontinuity of the $xx$
component of the stress tensor across the plate (which follows immediately
from the physical meaning of the stress tensor in terms of the flux of
momentum):
\be
F=\langle T_{xx}\rangle\bigg|_{x=a-}-\langle T_{xx}\rangle\bigg|_{x=a+}
=-\frac\pi{24a^2}.
\ee
Since energies are undefined up to a constant, without any loss of generality
we may take the stress tensor to be completely finite:
\be
\langle T^{\mu\nu}(x)\rangle\to\left\{\begin{array}{cc}
-\frac\pi{24a^2}\left(\begin{array}{cc}
1&0\\
0&1\end{array}\right),&0\le x\le a,\\
0,&x<0\quad\mbox{or}\quad x>a.
\end{array}\right.
\ee
\subsection{Massless Scalar in 3+1 Dimensions}
In higher dimensions, surface divergences appear.  These were discussed
in detail in Ref.~\cite{miltonbook}, \S 11.1, but for the sake of completeness
we repeat the discussion here.

In three space dimensions,
the use of the canonical stress tensor (\ref{stresstensor}) 
leads to the following expression for the vacuum
expectation value of the energy density,
\be
\langle T^{00}\rangle=\int\frac{d\omega}{2\pi}\frac{d^2k}{(2\pi)^2}\langle
t^{00}\rangle,\ee
where we have Fourier transformed both in frequency and transverse momentum.
If we take the plates to be located at $z=0$ and at $z=a$, we obtain
$\langle t^{00}\rangle$ by applying
the differential operator
\be
\frac12(\partial_0\partial_0'+\partial_x\partial_x'+\partial_y\partial_y'
+\partial_z\partial_z')\to\frac12(\omega^2+k^2+\partial_z\partial_z')
\ee
to the Green's function (\ref{ingf}) with $\omega\to\lambda\equiv
\sqrt{\omega^2-k^2}$,
\be
0\le x,x'\le a:\quad g(x,x';\lambda)=-\frac{\sin\lambda x_<\sin\lambda(x_>-a)}
{\lambda\sin\lambda a}.
\label{ingf2}
\ee
The result,
\begin{equation}
\langle t^{00}\rangle =- \frac{1}{2i\lambda\sin\lambda a}[\omega^2\cos\lambda a
-k^2\cos\lambda(2z-a)],
\end{equation}
is evaluated by making a Euclidean rotation,
\begin{equation}
\omega\to i\zeta,\quad \lambda\to i\kappa,
\end{equation}
and introducing polar coordinates in the $\zeta$, $k$
plane,
\begin{equation}
\zeta=\kappa\cos\theta,\quad k=\kappa\sin\theta,
\label{polarcoord}
\end{equation}
so
\begin{eqnarray}
\langle T^{00}\rangle(z)
&=&-\frac{1}{4\pi^2}\int_0^\infty \kappa\,d\kappa\int_0^{\pi/2}d\theta\,\kappa^2
\frac{\sin\theta}{\sinh\kappa a}[\cos^2\theta\cosh\kappa a\nonumber\\
&&\qquad\quad\qquad\qquad\qquad\mbox{}+\sin^2\theta\cosh\kappa
(2z-a)]\nonumber\\
&=&-\frac{1}{12\pi^2}\int_0^\infty d\kappa\,\kappa^3\frac{1}{\sinh\kappa a}
[\cosh\kappa a+2\cosh\kappa(2z-a)]\nonumber\\
&=&-\frac{1}{6\pi^2}\int_0^\infty d\kappa\,\kappa^3\left(\frac{1}
{e^{2\kappa a}-1}+
\frac{1}{2}+\frac{e^{2\kappa z}+e^{2\kappa(a-z)}}{e^{2\kappa a}-1}
\right).
\end{eqnarray}
Notice that the second term in the last integrand here corresponds to a constant
energy density, independent of $a$, so as before
it may be discarded as irrelevant. If we integrate the third term over $z$,
\begin{equation}
\int_0^a dz\left[e^{2\kappa z}+e^{2\kappa(a-z)}\right]=\frac{1}{\kappa}
\left[e^{2\kappa a}-1\right],
\end{equation}
we obtain another (divergent) constant term,
 so the only part of the vacuum
energy corresponding to an observable force is that coming from the first term:
\begin{equation}
\int_0^a dz\,\langle T^{00}\rangle(z)
=-\frac{a}{6\pi^2}\int_0^\infty d\kappa\frac{\kappa^3}{ e^{2\kappa
a}-1}=-\frac{\pi^2}{1440 a^3},
\label{localcasenergy}
\end{equation}
which is the well-known Casimir energy/area for a massless scalar field subject
to Dirichlet boundary conditions, one-half that for an electromagnetic
field \cite{casimir}.

In general, we have
\begin{subequations}
\begin{eqnarray}
\langle T^{00}\rangle (z)&=&u+g(z),\label{localed1}
\end{eqnarray}
where
\begin{eqnarray}
u&=&-\frac{\pi^2}{1440 a^4},
\label{localed}\\
g(z)&=&-\frac{1}{6\pi^2}\frac{1}{16a^4}\int_0^\infty dy\,y^3\frac{e^{yz/a}
+e^{y(1-z/a)}}{ e^y-1}.
\end{eqnarray} 
\end{subequations} 
If we expand the denominator in a geometric series,
\begin{equation}
\frac{1}{e^y-1}=\frac{e^{-y}}{1-e^{-y}}=\sum_{n=1}^\infty e^{-ny}, 
\end{equation}
we can express $g$ in terms of the generalized or Hurwitz zeta 
function,
\begin{equation}
\zeta(s,a)\equiv \sum_{n=0}^\infty \frac{1}{(n+a)^s}, \quad a\ne \mbox{ a 
negative integer},
\end{equation}
as follows:
\begin{equation}
g(z)=-\frac{1}{16\pi^2 a^4}[\zeta(4,z/a)+\zeta(4,1-z/a)].
\label{gzzeta}
\end{equation}
This function is plotted in Fig.~\ref{fig:local1}, where it will be
observed that it diverges quartically as $z\to 0$, $a$.  (Its $z$ integral
over the region between the plates diverges cubically.)  As we have seen,
this badly behaved function does not contribute to the force on the
plates.

\begin{figure}
\centerline{
\psfig{figure=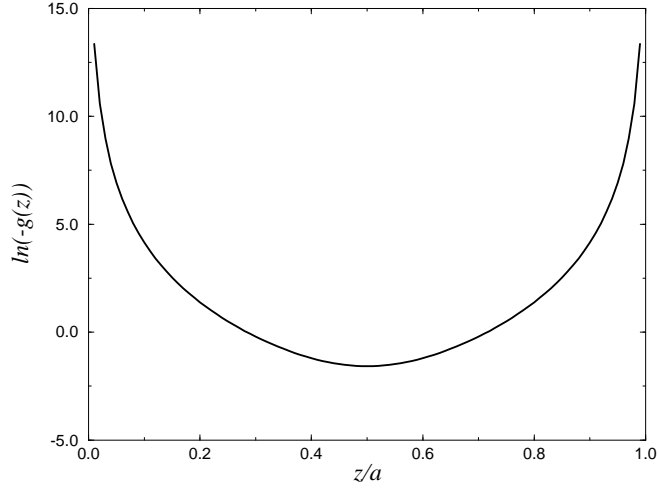,width=3in,angle=270}}
\caption{The singular part of the local energy density between
parallel plates at $z=0$ and $z=a$.}
\label{fig:local1}
\end{figure}

Next, we turn to $\langle T_{zz}\rangle$.  According to the stress tensor
(\ref{stresstensor}) and the Green's function (\ref{ingf2}),
that is given by
\begin{eqnarray}
\langle T_{zz}\rangle&=&\frac{1}{2i}(\partial_z\partial'_{z}-\partial_x
\partial'_{x}-\partial_y\partial'_{y}+\partial_0\partial'_{0})G(x,x')
\nonumber\\
&=&\frac{1}{2i}\int\frac{d\omega\, d^2k}{(2\pi)^3}(\partial_z\partial'_{z}
+\lambda^2)\left[-\frac{1}{\lambda\sin\lambda a}\sin\lambda z_<\sin\lambda
(z_>-a)\right]\nonumber\\
&=&-\frac{1}{2i}\int\frac{d\omega \,d^2k}{(2\pi)^3}\frac{\lambda}{\sin\lambda a}
[\cos\lambda z\cos\lambda(z-a)+\sin\lambda z\sin\lambda(z-a)]\nonumber\\
&=&\int\frac{d\omega \,d^2k}{(2\pi)^3}\frac{i\lambda}{2}\cot\lambda a,
\end{eqnarray}
which is independent of $z$; that is, the 
normal-normal component of the 
expectation value of the stress tensor between the plates is constant.
If once again, the irrelevant $a$-independent part is removed,\footnote{The
infinite parts of $\langle T_{00}\rangle$ and $\langle T_{zz}\rangle$ 
are related by the same factor of three as the finite parts:
\begin{equation}
\langle T_{zz}\rangle^{\rm inf}=-\int\frac{d\kappa\,\kappa^3}{4\pi^2},\quad
\langle T_{00}\rangle^{\rm inf}=-\int\frac{d\kappa\,\kappa^3}{12\pi^2}.
\end{equation}}
what is left 
is just three times the constant part of the energy density (\ref{localed}),
\begin{equation}
\langle T_{zz}\rangle =-3\times \frac{\pi^2}{1440 a^4}.
\end{equation}

The remaining nonzero components of the stress tensor are
\begin{eqnarray}
\langle T_{xx}\rangle&=&\langle T_{yy}\rangle=\frac{1}{2i}[
\partial_x\partial'_{x} -\partial_y\partial'_{y}-\partial_z\partial'_{z}+
\partial_0\partial'_{0}]G(x,x')\nonumber\\
&=&-\frac{1}{2i}\int\frac{d\omega\,d^2k}{(2\pi)^3}\frac{1}{\lambda\sin\lambda a}
[\omega^2\sin\lambda z\sin\lambda(z-a)\nonumber\\
&&\qquad\qquad\quad\qquad\qquad\mbox{}-\lambda^2\cos\lambda z\cos\lambda(z-a)]
\nonumber\\
&=&-u-g(z),
\label{localtxx}
\end{eqnarray}
where we have again introduced polar coordinates in the frequency-wavenumber
plane, and again dropped the infinite ($a$-independent) constant in $u$. 
Thus the tensor structure of stress tensor is
\begin{equation}
\langle T^{\mu\nu}\rangle(z)=u\left(\begin{array}{cccc}
1&0&0&0\\
0&-1&0&0\\
0&0&-1&0\\
0&0&0&3
\end{array}\right)+g(z)\left(\begin{array}{cccc}
1&0&0&0\\
0&-1&0&0\\
0&0&-1&0\\
0&0&0&0
\end{array}\right),
\label{tmunucan}
\end{equation}
where $u$ is given by (\ref{localed}) and $g$ by (\ref{gzzeta}).
Because $u$ is constant, this vacuum expectation value is divergenceless,
since $g(z)$ does not contribute to $\langle T^{zz}\rangle$:
\begin{equation}
\partial_\mu \langle T^{\mu\nu}\rangle=\partial_z\langle  T^{zz}\rangle=0.
\end{equation}
The second term in (\ref{tmunucan}) diverges at the boundaries, $z=0$, $a$,
and has a integral over the volume which diverges; yet as we have seen, it is
physically irrelevant because its integral is independent of 
$a$, and it has no normal
component.  Is there a natural way in which it simply does not appear in the
local formulation?

The affirmative answer hinges on the ambiguity in defining the stress 
tensor.\footnote{For a rather complete discussion of this see 
Ref.~\cite{schwingerpsf}, Secs.~3-7, 3-17.}
It was noted in Ref.~\cite{miltonbook} 
that this ambiguity was without effect as far as the
total stress or the total energy was concerned.\index{Stress tensor!ambiguity} 
 Now, however, we see the
virtue of the conformal stress tensor 
\cite{ccj}:
\begin{equation}
\tilde T^{\mu\nu}=\partial^\mu\phi\partial^\nu\phi-\frac{1}{2}g^{\mu\nu}
\partial_\lambda\phi\partial^\lambda\phi-\frac{1}{6}(\partial^\mu\partial^\nu
-g^{\mu\nu}\partial^2)\phi^2,
\end{equation}
which, because of the equation of motion $\partial^2\phi=0$, has a vanishing
trace, 
\begin{equation}
\tilde T^\mu{}_\mu=0.
\end{equation}  
If we use this stress tensor rather than the canonical one, we merely need
supplement the above computations by that of the vacuum expectation value
of the extra term.  Thus to obtain $\langle \tilde T^{xx}\rangle$ we add to
(\ref{localtxx})
\begin{eqnarray}
&&\frac{1}{6i}(\partial_y^2+\partial_z^2-\partial_0^2)G(x,x)\nonumber\\
&=&\frac{1}{6i}\int\frac{d\omega\,d^2k}{(2\pi)^3}\partial_z^2\left[-
\frac{1}{\lambda\sin\lambda a}\sin\lambda z\sin\lambda(z-a)\right]\nonumber\\
&=&-\frac{1}{6i}\int\frac{d\omega\,d^2k}{(2\pi)^3}\frac{2\lambda}{\sin\lambda z}
\cos\lambda(2z-a)\nonumber\\
&=&g(z), 
\label{extraconformal}
\end{eqnarray}
which just cancels the surface-divergent
 term in (\ref{localtxx}). Again, because $G(x,x)$
only depends on $z$, there is no extra contribution to $\langle T_{zz}\rangle$:
\begin{equation}
-\frac{1}{6}(\partial_z^2-g_{zz}\partial^2)\langle\phi^2\rangle
=\frac{1}{6i}(\partial_x^2+\partial_y^2-\partial_0^2)G(x,x)=0.
\end{equation}
The extra term for $\langle T_{00}\rangle$ is just the negative of that
in (\ref{extraconformal}),
\begin{equation}
-\frac{1}{6i}\partial_z^2G(x,x)=-g(z),
\end{equation}
which cancels the second term in (\ref{localed1}).  Thus, the conformal
stress tensor has the following vacuum expectation value for the
region between the parallel plates:
\begin{equation}
\langle \tilde T^{\mu\nu}\rangle =u\left(\begin{array}{cccc}
1&0&0&0\\
0&-1&0&0\\
0&0&-1&0\\
0&0&0&3
\end{array}\right)
\label{st-pp}
\end{equation}
which is traceless, thereby respecting the conformal invariance of the massless
theory.  This is just the result found by Brown and Maclay by general
considerations \cite{brown}, who argued that
\begin{equation}
\langle \tilde T^{\mu\nu}\rangle=u[4\hat z^\mu\hat z^\nu-g^{\mu\nu}], 
\end{equation}
where $\hat z^\mu$ is the unit vector in the $z$ direction.

\subsection{Massive Scalar in $D$ Spatial Dimensions}
It is instructive to repeat the above calculation for a massive scalar
field where the plates have $D-1$ transverse dimensions.  We will use
the conformal stress tensor,
\be
T^{\mu\nu}=\partial^\mu\phi\partial^\nu\phi-\frac12g^{\mu\nu}
(\partial_\lambda\phi
\partial^\lambda\phi+\mu^2\phi^2)-\alpha(\partial^\mu\partial^\nu-g^{\mu\nu}
\partial^2)\phi^2.
\ee
Here $\alpha$ has to be chosen to be $(D-1)/(4D)$ in order that the trace
vanish (by virtue of the field equations) in the massless limit:
\be
\alpha=\frac{D-1}{4D}:\quad T^\mu{}_\mu=-\mu^2\phi^2.
\ee
The calculation proceeds very similarly to that given above.  The only
new element is writing the momentum integral in polar coordinates:
\be
d^{D-1}k=\frac{2\pi^{(D-1)/2}}{\Gamma\left(\frac{D-1}{2}\right)}k^{D-2}\,dk,
\ee
and then introducing polar coordinates as in Eq.~(\ref{polarcoord}).
We encounter the integrals
\be
\int_0^{\pi/2} d\theta\,(\sin\theta)^{D-2}=2^{D-3}\frac{\Gamma\left(
\frac{D-1}2\right)^2}{\Gamma(D-1)},
\ee
relative to which
\be
\langle\sin^2\theta\rangle=\frac{D-1}D,\quad\langle\cos^2\theta\rangle=\frac1D.
\ee
The result for the various nonzero components of the stress tensor are
($\kappa^2=\rho^2+\mu^2$)
\begin{subequations}
\bea
\langle T^{00}\rangle&=&-\frac{2^{-D}\pi^{-D/2}}{D\Gamma(D/2)}
\int_0^\infty d\rho\,\rho^{D-1}\frac1{\kappa\sinh\kappa a}[\rho^2\cosh\kappa a
+\mu^2\cosh\kappa(2z-a)],\\
\langle T^{zz}\rangle&=&-\frac{2^{-D}\pi^{-D/2}}{\Gamma(D/2)}
\int_0^\infty d\rho\,\rho^{D-1}\kappa\coth\kappa a,
\\
\langle T^{xx}\rangle&=&\langle T^{yy}\rangle=\dots=-\langle T^{00}\rangle.
\eea
\end{subequations}
Surface divergent terms, which do not contribute to the observable force,
appear proportional to the square of the mass.
Of course, the trace of the expectation value of the stress tensor is
nonzero because of the mass:
\bea
\langle T^\mu{}_\mu\rangle&=&\langle T^{zz}\rangle-D\langle T^{00}\rangle
=-\mu^2\langle\phi^2\rangle\nonumber\\
&=&-\mu^2\frac{2^{-D}\pi^{-D/2}}{\Gamma(D/2)}
\int_0^\infty d\rho\,\rho^{D-1}\frac1{\kappa\sinh\kappa a}[\cosh\kappa a
-\cosh\kappa(2z-a)].
\eea
As before, the infinite $a$-independent stress which would be present
if the boundary were not present ($\coth\kappa a\to1$) may be
removed, as it does not contribute to the force on the plates.  
The well known \cite{ambjorn} expressions for the force and the
energy between parallel plates may be easily recovered.
We do not see the necessity for the additional terms found by Graham
et al. \cite{graham2} in the energy to make the energy finite at zero 
separation (the fact that the Casimir energy diverges at $a=0$ reflects the
infinite amount of energy released when the plates are pushed into
coincidence) nor the requirement that the energy should be infinite
at zero mass, when the observable force is finite there.

\section{Scalar Casimir Effect for a Dirichlet Sphere}
\label{Sec3}

The calculation given in Sec.~\ref{Sec2.1} was that for a sphere in one
spatial dimension.  Now we consider a massless scalar in three space
dimensions, with a spherical boundary on which the field vanishes.
This corresponds to the TE modes for the electrodynamic situation first
solved by Boyer \cite{boyersphere}.  The general calculation in $D$
dimensions was given in Ref.~\cite{benmil}; the force per unit area is
given by the formula
\be
{\cal F}=-\sum_{l=0}^\infty\frac{(2l+D-2)\Gamma(l+D-2)}{l!2^D\pi^{(D+1)/2}
\Gamma(\frac{D-1}2)a^{D+1}}\int_0^\infty dx\,x\frac{d}{dx}\ln\left[I_\nu(x)
K_\nu(x)x^{2-D}\right].\ee
Here $\nu=l-1+D/2$.  For $D=3$ this expression reduces to
\be
{\cal F}=-\frac1{8\pi^2a^4}\sum_{l=0}^\infty(2l+1)\int_0^\infty dx\,x\frac{d}
{dx}\ln\left[I_{l+1/2}(x)K_{l+1/2}(x)/x\right].
\label{fsphere}
\ee
In Ref.~\cite{benmil} we evaluated this expression by continuing in $D$ from
a region where both the sum and integrals existed.  In that way, a completely
finite result was found for all positive $D$ not equal to an even integer.

Here we will adopt a perhaps more physical approach, that of allowing the 
time-coordinates in the underlying Green's function to approach each other,
as described in Ref.~\cite{mildersch}.  That is, we recognize that the $x$
integration above is actually a (dimensionless) frequency integral, and
therefore we should replace
\be
\int_0^\infty dx\,f(x)=\frac12\int_{-\infty}^\infty dy\,e^{iy\delta}f(|y|),
\ee
where at the end we are to take $\delta\to0$.  Immediately, we can
replace the $x^{-1}$ inside the logarithm in Eq.~(\ref{fsphere})
 by $x$, which makes the integrals
converge, because the difference is proportional to a delta function in
the time separation, a contact term without physical significance.

To proceed, we use the uniform asymptotic expansions for the modified
Bessel functions, as described in detail in Ref.~\cite{miltonbook}.  This
is an expansion in inverse powers of $\nu=l+1/2$, low terms in which turn
out to be remarkably accurate even for modest $l$.  The leading terms
in this expansion are
\be
\ln\left[x I_{l+1/2}(x)K_{l+1/2}(x)\right]
\sim\ln\frac{zt}2+\frac1{\nu^2}g(t)+\frac1{\nu^4}
h(t)+\dots,
\label{uae}
\ee
where $x=\nu z$ and $t=(1+z^2)^{-1/2}$.
Here
\begin{subequations}
\bea
g(t)&=&\frac18(t^2-6t^4+5t^6),\\
h(t)&=&\frac1{64}(13t^4-284t^6+1062t^8-1356t^{10}+565t^{12}).
\eea
\end{subequations}
The leading term in the force/area is therefore
\bea
{\cal F}_0&=&
-\frac1{8\pi^2a^4}\sum_{l=0}^\infty(2l+1)\nu\int_0^\infty dz\, t^2\nonumber\\
&=&-\frac1{8\pi a^4}\sum_{l=0}^\infty\nu^2=\frac3{32\pi a^4}\zeta(-2)=0.
\eea
where in the last step we have used a formal zeta function 
evaluation.\footnote{Note that the corresponding TE contribution for
the electromagnetic Casimir effect would not be zero, for there the sum
starts from $l=1$.}  Here the rigorous way to argue is to recall the
presence of the point-splitting factor $e^{i\nu z\delta}$ and to carry out
the sum on $l$ using
\be
\sum_{l=0}^\infty e^{i\nu z\delta}=-\frac1{2i}\frac1{\sin z\delta/2},
\ee
so
\bea
\sum_{l=0}^\infty \nu^2e^{i\nu z\delta}&=&-\frac{d^2}{d(z\delta)^2}\frac{i}
{2\sin z\delta/2}\nonumber\\
&=&\frac{i}8\left(-\frac2{\sin^3z\delta/2}+\frac1{\sin z\delta/2}\right).
\eea
Then ${\cal F}_0$ is given by the divergent expression
\be
{\cal F}_0=\frac{i}{\pi^2 a^4\delta^3}\int_{-\infty}^\infty \frac{dz}{z^3}
\frac1{1+z^2},
\ee
which we argue is zero because the integrand is odd.

The next term in the uniform asymptotic expansion (\ref{uae}), 
that involving $g$, 
likewise gives zero pressure, as intimated by the formal zeta function identity,
\be
\sum_{l=0}^\infty \nu^s=(2^{-s}-1)\zeta(-s),
\ee
which vanishes at $s=0$.  The same conclusion follows from point splitting,
as we can see through use of the Euler-Maclaurin sum formula,
\be
\sum_{l=0}^\infty f(l)=\int_0^\infty dl\,f(l)+\frac12f(0)-\sum_{k=1}^\infty
\frac{B_k}{(2k)!}f^{(2k-1)}(0).
\ee
Here we have
\be
\int_0^\infty dl\,e^{i\nu z\delta}=-\frac{e^{iz\delta/2}}{iz\delta}
=-\frac1{i z\delta}-\frac12+{\cal O}(\delta).
\ee
We argue again that the first term here gives no contribution to the integral
over $z$ because it is odd, 
and then the first two terms in the Euler-Maclaurin formula give
\be
{\cal F}_1=-\frac1{8\pi^2a^4}\left[-\frac12\int_{-\infty}^\infty dz\,z\frac{d}{dz}
g(t)+\frac12\int_{-\infty}^\infty dz\,z\frac{d}{dz}g(t)\right]=0.
\ee
Derivatives of $e^{i\nu z\delta}$ with respect to $l$ all vanish at $z=0$.
  Again, this cancellation does not occur in the electromagnetic case
because there the sum starts at $l=1$.

So here the leading term which survives is that of order $\nu^{-4}$
in Eq.~(\ref{uae}),
namely
\be
{\cal F}_2=\frac1{4\pi^2a^4}\sum_{l=0}^\infty \frac1{\nu^2}\int_0^\infty
dz \,h(t),
\ee
where we have now dropped the point splitting factor because this expression
is completely convergent.  The integral over $z$ is
\be
\int_0^\infty dz \, h(t)=\frac{35\pi}{32768}
\ee and the sum over $l$ is $3\zeta(2)=\pi^2/2$, so the leading contribution
to the stress on the sphere is
\be
{\cal S}_2=4\pi a^2{\cal F}_2=\frac{35\pi^2}{65536a^2}=\frac{0.00527094}{a^2}.
\ee
Numerically this is a terrible approximation.

What we must do now is return to the full expression and add and subtract
the leading asymptotic terms.  This gives
\be
{\cal S}={\cal S}_2-\frac1{2\pi a^2}\sum_{l=0}^\infty(2l+1)R_l,
\ee
where 
\be
R_l=Q_l+\int_0^\infty dx\left[\ln zt+\frac1{\nu^2}g(t)+\frac1{\nu^4}h(t)\right],
\label{remainder}
\ee
where the integral
\be
Q_l=-\int_0^\infty dx\ln[2xI_\nu(x)K_{\nu}(x)]
\ee
was given the asymptotic form  in Ref.~\cite{benmil}
\bea
Q_l&\sim&\frac{\nu\pi}2+\frac\pi{128\nu}-\frac{35\pi}{32768\nu^3}
+\frac{565\pi}{1048577\nu^5}\nonumber\\
&&\quad\mbox{}-\frac{1208767\pi}{2147483648\nu^7}
+\frac{138008357\pi}{137438953472\nu^9},\quad l\gg1.
\label{ql}
\eea
The first two terms in Eq.~(\ref{ql}) cancel the second and third terms in 
Eq.~(\ref{remainder}), of course.
The third term in Eq.~(\ref{ql}) corresponds to $h(t)$, so the last three terms 
displayed in Eq.~(\ref{ql}) give the asymptotic behavior of the remainder,
which we call $w(\nu)$.  Then we have, approximately,
\be
{\cal S}\approx {\cal  S}_2-\frac1{\pi a^2}\sum_{l=0}^n\nu R_l-\frac1{\pi a^2}
\sum_{l=n+1}^\infty \nu w(\nu).
\ee
For $n=1$ this gives ${\cal S}\approx0.00285278/a^2$, and for larger $n$
this rapidly approaches the value first  given in Ref.~\cite{benmil}:
\be
{\cal S}=0.002817/a^2,
\ee
a value much smaller than the famous electromagnetic result \cite{boyersphere,
davis,mildersch,balian},
\be
{\cal S}^{\rm EM}=\frac{0.04618}{a^2},
\ee
because of the cancellation of the leading terms noted above.

\newcommand{\Tr}{\mbox{Tr}\,}

\section{Diagrammatic Divergence Structure}
\label{Sec4}
In the two previous sections we have come to rather different conclusions from those
of Ref.~\cite{graham2}.  For the case of parallel plates, studied in 
Sec.~\ref{Sec2} we found:
\begin{itemize}
\item The massless theory is perfectly well defined (no infrared divergences), and
surface divergences, which in any case have no physical consequences, do not
appear if the conformal stress tensor is used.
\item The vacuum expectation value of the stress tensor for the case of a massive
scalar does have surface divergences, which are proportional to the mass squared, but
which do not contribute to the force and are therefore physically irrelevant.
\end{itemize}
For a massless scalar with a spherical boundary in three dimensions, the formal
expressions for the force/area and the energy are formally divergent, yet if they
are regulated, say by point-splitting, the divergences cancel and the energy and
self-stress on the sphere are completely finite and unambiguous.

The authors in Ref.~\cite{graham2} came to different
 conclusions.  However, their disagreement
with us on the $D=1$ case seems entirely semantic, and without observable consequence.
Their substantial argument hinged on their $D=2$ calculation.  However, it is well known
that the Casimir effect for a circle is divergent, so it is hard to draw general
inferences from an examination of that situation.  Here, we will re-examine some of the
general arguments of Ref.~\cite{graham2} for a hypersphere in $D$ space dimensions.

The general analysis for that case was given in Ref.~\cite{benmil}; it is clear that
the point-splitting method given in the previous section could be applied in 
that calculation.
Instead, we will here focus on the issue of the second-order Feynman graph which
supposedly is the signal for the divergence of the theory in any number of 
space dimensions.  (It is the oversubtracted graph which leaves the
mode sum more convergent.)
We will adopt a somewhat simpler formalism than that given in Ref.~\cite{graham2}, based
on the ``trace-log'' formula for the energy,
\begin{subequations}
\be
E=\frac{i}{2T}\Tr \ln G,
\label{trln}
\ee
where for a ``polarization'' operator $\Pi$
\be
G=G_0(1+\Pi G)=G_0(1+\Pi G_0+\Pi G_0\Pi G_0+\dots).
\label{gpig}
\ee
\end{subequations}

The highly sensible approach of Graham et al.~\cite{graham,graham2} is to replace
ideal boundary conditions by an interaction with an external field $\sigma$.
The Lagrangian for the scalar field is thus taken to be
\be
{\cal L}=-\frac12(\partial_\mu\phi\partial^\mu\phi+m^2\phi^2+\sigma(r)\phi^2),
\ee
where, anticipating spherical symmetry, we have taken the external field to depend
only on the spatial radial coordinate.  In the end, we may take $\sigma$ to
be a delta function,
\be
\sigma(r)=\frac{g}{a}\delta(r-a),
\ee
where $g$ is dimensionless
and the formal $g\to\infty$ limit corresponds to the situation of a 
Dirichlet
spherical shell.  We can now evaluate the one-loop vacuum energy by the replacement
$\Pi\to\sigma$ in Eqs.~(\ref{trln}), (\ref{gpig}).  It is the second-order graph
that is supposed to signal nonrenormalizability. 
\subsection{General $D$}
We first carry out the calculation in $D$ dimensions.
\bea
E&=&\frac{i}{2T}\Tr\sigma G_0\sigma G_0\nonumber\\
&=&\frac{i}{2T}\int d^{D+1}x\,d^{D+1}y\,
\sigma(x)G_0(x-y)\sigma(y)G_0(y-x)\nonumber\\ &=&\pi i\int d^Dx \,d^Dy\,
\sigma(|x|)\sigma(|y|)\int\frac{d\omega}{2\pi}\int\frac{d^Dp}{(2\pi)^D}
\frac{d^Dq}{(2\pi)^D}\frac{e^{i{\bf (p-q)\cdot(x-y)}}}{(p^2+m^2)(q^2+m^2)},
\eea
where in the last line we have carried out the integral on $t$ and $t'$, and as a result
$p^0=q^0=\omega$.  Now we introduce polar coordinates, so in terms of the
last angle
\be
d^D x=A_{D-1}x^{D-1}dx \sin^{D-2}\theta\, d\theta,
\ee
where $A_n=2\pi^{n/2}/\Gamma(n/2)$ is the surface 
area of a sphere in $n$ dimensions.  Then we encounter a Bessel function
\be
\int_0^\pi d\theta\sin^{D-2}\theta\, e^{i|{\bf p-q}|x\cos\theta}
=\left(\frac2{|{\bf p-q}|x}\right)^{D/2-1}\sqrt{\pi}\,
\Gamma\left(\frac{D-1}2\right)
J_{D/2-1}(|{\bf p-q}|x).
\ee
Thus the Fourier transform of the field $\sigma(|x|)$ is defined by
\bea
\tilde\sigma(k)&=&\int d^D x \,e^{i{\bf k\cdot x}}\,\sigma(x)\nonumber\\
&=&k\left(\frac{2\pi}k\right)^{D/2}\int_0^\infty dx\,x^{D/2}J_{D/2-1}(kx)
\sigma(x).\label{4.7}
\eea
(This agrees with the expression in Ref.~\cite{graham2} for $D=2$.)

The expression for the energy reduces to
\be
E=i\pi\int\frac{d\omega}{2\pi}\int\frac{d^Dq \, d^Dp}{(2\pi)^{2D}}
\frac1{(p^2+m^2)
(q^2+m^2)}
\bigg|_{p^0=q^0=\omega}\tilde\sigma(|{\bf p-q}|)^2.
\ee
We carry out the momentum integrations by first using the proper-time 
representation to combine the denominators:
\bea
\frac1{p^2+m^2}\frac1{q^2+m^2}&=&\int_0^\infty ds\int_0^\infty ds' 
e^{-s(p^2+m^2)-s'(q^2+m^2)}\nonumber\\
&=&\int_0^\infty ds\,s\int_0^1 du\,e^{-sm^2-s(1-u)p^2-suq^2},
\eea
where in the second line we replace $s\to s(1-u)$, $s'\to su$.  In terms of
$\bf k=p-q$, we complete the square in the exponent by writing
\be
s(1-u){\bf p}^2+su{\bf q}^2=s[({\bf p}-u{\bf k})^2+{\bf k}^2u(1-u)],
\ee
while the corresponding $0$ components combine to give $-s\omega^2$.  Now the
frequency and ${\bf p}$ integrals are just Gaussian:
\be
\int d\omega \,e^{s\omega^2}=i\sqrt{\frac\pi s},\quad
\int d^D(p-uk)\,e^{-s({\bf p}-u{\bf k})^2}=\left(\frac\pi s\right)^{D/2}.
\ee
Finally, we introduce polar coordinates for the $\bf k$ integration, with the result
\be
E=-2^{-2D}\pi^{-D+1/2}\frac{\Gamma\left(\frac{3-D}{2}\right)}
{\Gamma\left(\frac{D}{2}\right)}\int_0^\infty dk\,k^{D-1}\tilde
\sigma(k)^2\int_0^1du\,[m^2+u(1-u)k^2]^{D/2-3/2},
\label{4.12}
\ee
which yields the $D=2$ result given in Ref.~\cite{graham2}.

If we choose a delta-function potential,
\be
\sigma(x)=\frac{g}{a}\delta(x-a)
\ee
we obtain
\be
E=-2^{-D}\pi^{1/2}\frac{\Gamma\left(\frac{3-D}2\right)}{\Gamma\left(\frac{D}2
\right)}\frac{g^2}{a}\int_0^\infty d\xi\,\xi\,J_{D/2-1}^2(\xi)\int_0^1du\,[m^2a^2
+\xi^2u(1-u)]^{(D-3)/2}.
\ee
This appears to converge for $0<D<2$ 
except for the exceptional case $m=0$.  In that
case the $u$ integral is simply
\be
\frac{\Gamma\left(\frac{D-1}2\right)^2}{\Gamma(D-1)}
=2^{2-D}\pi^{1/2}\frac{\Gamma\left(\frac{D-1}2\right)}{\Gamma(\frac{D}2)},
\ee and the integral over the Bessel functions is
\bea
\int_0^\infty d\xi\, \xi^{D-2}J^2_{D/2-1}(\xi)&=&2^{D-2}\frac{\Gamma(2-D)
\Gamma(D-3/2)}{\Gamma\left(\frac{3-D}2\right)^2\Gamma(\frac12)}\nonumber\\
&=&\frac{\Gamma(1-D/2)\Gamma(D-3/2)}{2\pi\Gamma\left(\frac{3-D}2\right)},
\eea
which is valid in the region
\be
\frac32<D<2.
\ee
Thus the energy for a massless scalar is
\be
E=-2^{1-2D}\frac{g^2}{a}\frac{\Gamma\left(\frac{D-1}2\right)\Gamma(D-3/2)
\Gamma(1-D/2)}{\Gamma\left(\frac{D}2\right)^2},
\label{result}
\ee
which we take to be the appropriate analytic continuation for all $D$.
This exhibits poles at $D=2, 4, 6, \dots$, in congruence with the known
divergence structure of the Casimir effect.  There are also poles occurring
at $D=1, -1, -3, \dots$, and at $D=3/2, 1/2, -1/2, \dots$.  These latter
two sequences of divergent dimensions correspond to infrared divergences
that have no counterpart in the Casimir calculations, unlike the ultraviolet,
even-integer poles. For space dimension between 2 and 4 the Casimir energy
is completely finite, in concert with this diagnostic.  The divergence at
$D=2$, even putting aside the question of mass, is seen not to be generic.

\subsection{$D=3$}
Instead of dimensional continuation, one can work directly in $D=3$.
Let us regulate the theory by inserting a lower limit $s_0\to 0$ in the
proper-time integration, so that for $m=0$ the energy (\ref{4.12}) becomes
\bea
E&=&\frac{1}{2^5\pi^3}\int_0^\infty dk\,k^2\tilde\sigma(k)^2\int_0^1 du
\ln[s_0k^2u(1-u)]\nonumber\\
&=&
\frac{1}{2^5\pi^3}\int_0^\infty dk\,k^2\tilde\sigma(k)^2\frac{d}{d\alpha}
\int_0^1 du[s_0k^2u(1-u)]^\alpha\bigg|_{\alpha=0}.
\eea
\normalcolor
If the derivative acts on anything but $k^{2\alpha}$ we have
\be
\int_0^\infty dk\,k^2\tilde\sigma(k)^2=(2\pi)^3\int_0^\infty dx\,x^2\sigma(x)^2.
\ee
This diverges as $\sigma(x)\to(g/a)\delta(x-a)$;
but if we regulate the divergence by point-splitting
\be
\sigma(x)^2\to\lim_{\xi\to\infty}\sigma(x-\xi)\sigma(x+\xi),
\ee
we have
\be
\int_0^\infty dk\,k^2\tilde\sigma(k)^2=(2\pi)^3g^2\delta(2\xi),
\ee
which is seen to be a contact term, independent of $a$.

We are left with
\bea
E&=&\frac{1}{2^5\pi^3}\frac{d}{d\alpha}\int_0^\infty dk\,k^2\tilde\sigma(k)^2
k^{2\alpha}\nonumber\\
&=&\frac1{2\pi}\int_0^\infty dx\,x\,\sigma(x)\int_0^\infty dy \,y\,
\sigma(y)\frac{d}{d\alpha}\int_0^\infty dk\,k^{2\alpha}\sin kx\sin ky
\bigg|_{\alpha=0}.
\eea
Here we have used the Fourier transformation expression  (\ref{4.7}), but
replaced Bessel functions of order 1/2 by the corresponding trigonometric
functions.  The $k$ integral is now obtained from ($-1<\alpha<0$)
\bea
\int_0^\infty dx\,x^\alpha\cos\beta x=\frac{\Gamma(\alpha+1)\cos(\alpha+1)
\frac\pi2}{\beta^{\alpha+1}},
\eea
so that the energy becomes
\bea
E=\frac1{4}\int_0^\infty dx\,x\sigma(x)\int_0^\infty dy \,y\sigma(y)
\left(\frac1{x+y}-\frac1{|x-y|}\right)\to\frac{g^2}{8a},
\label{d=3energy}
\eea
where we have omitted another infinite term that is independent of $a$.
The result is exactly the $D=3$ value of Eq.~(\ref{result}).
The justification for omitting (infinite) constant terms
in the energy is that they are unobservable, not corresponding to a 
self-stress on the sphere.  See also the Appendix.

We further might observe that this energy (\ref{d=3energy})
could not be rendered finite were a Gaussian
profile rather than a delta function employed.  This is not surprising.
Finiteness is only anticipated for an infinitesimal shell, and not for
a smooth boundary with continuously changing properties.  For example,
the Casimir energy
for a thick dielectric shell apparently contain irremovable divergences.
\section{Conclusions}
\label{Sec5}
The challenge set forth in Ref.~\cite{graham} and elaborated in 
Refs.~\cite{graham2} is physically appropriate and timely given the
development of our understanding of the Casimir effect.  Certainly
those authors are justified in objecting to the loose use of 
the term ``renormalization'' in connection with various dubious processes
for removing divergences in boundary-value Casimir problems.  However,
it is important to separate the wheat from the chaff.  The Casimir
force between parallel plates, the self-stress (or the force per unit
area) on a perfect (Dirichlet or Neumann) spherical or cylindrical 
\cite{deraadcyl}
shell due to massless fields, the energy of fields confined to a curved
manifold (a hypersphere or  torus for example) 
\cite{appel,candandwein,nojiri,entropy}
are examples where the Casimir
energy is unambiguous and finite, except for exceptional numbers of spatial
dimensions.

Of course, these are special cases, and generically Casimir energies
are infinite.  This is true if fields bounded by a spherical shell have
mass, if the shell has finite thickness, or if
 the speed of light inside and outside the shell has different values.
The latter case is the interesting one of a dielectric ball, first
considered in Ref.~\cite{miltonballs}.  The stress or the energy in that
case is quartically divergent.  It was argued, very tentatively in
Ref.~\cite{miltonballs}, and more forcefully later
\cite{borelkirles}, that the divergent
terms could be reabsorbed into the definition of physical properties
of the material medium, the mass density, surface tension, and the like.
This ``renormalization'' was in the spirit of the first use of renormalization
in physics \cite{poisson,hlamb}. Obviously, this was not a very convincing
argument, and was not on a par with perturbative renormalization of
a quantum field theory.\footnote{The contrary opinion is expressed by
Ref.~\cite{bgnv}.}
  However, fairly recently, the discovery by
several groups \cite{bmm,barton,bkv,hb,lamb}
 that the finite part of the Casimir energy for a dilute
dielectric sphere was unique, and coincided with that obtained by a
regulated (dimensionally continued) calculation of the van der Waals
energy \cite{sonokm},
 did provide some evidence that the divergences could be removed
unambiguously, and had the practical consequence of destroying the hope
of explaining sonoluminescence on the basis of quantum vacuum energy
\cite{sonorev02}.

Obviously we are still at the early stages of understanding quantum
field theory.  The nature of divergences in vacuum energy calculations
is still not understood.  However, there are a few established peaks
that rise above the murky clouds of ignorance, and we should not abandon
them because the rest is obscure.

\appendix*
\section{Dimensional Continuation of Feynman Diagrams}
In spite of its impressive successes, one might have concern about the
use of dimensional continuation to evaluate divergent Feynman diagrams,
such as those considered in Sec.~\ref{Sec4}.  What is the meaning
of the dimensionally continued expression in dimensions where the
formula gives a finite result, even though the Feynman integral is
manifestly divergent?  Here we give a simple example of what is going
on.

Consider a $\lambda\phi^4$ theory in $d$ spacetime dimensions.  The lowest-order
self energy diagram gives
\be
\Sigma^{(1)}=-12\lambda\mu^{4-d}I_d,
\ee
where by the well known dimensional regularization formula (trivially derived
by simple proper-time manipulations)
\be
I_d=\int\frac{d^dl}{(2\pi)^d}\frac1{l^2+m^2}=\frac{m^{d-2}}{(4\pi)^{d/2}}
\Gamma\left(1-\frac{d}2\right).
\label{Id}
\ee
This equality is derived assuming $d<2$.  The right-hand side of this equation
diverges for even $d>2$.  Elementary field theory treatments of scalar field
theory have no hesitation in accepting that Eq.~(\ref{Id}) is valid in the
neighborhood of $d=4$.  Therefore, we should ask what does it mean for $d=3$?
There, the dimensionally continued formula says
\be
I_3=-\frac{m}{4\pi}.
\label{I3}
\ee

If we put in a large momentum cutoff $\Lambda$, we can compute the Feynman
integral directly:
\be
I_3=\frac{4\pi}{(2\pi)^3}\int_0^\Lambda l^2\,dl\frac1{l^2+m^2}=
\frac{\Lambda}{2\pi^2}-\frac{m}{4\pi}.
\ee
Of course, the integral is linearly divergent as $\Lambda\to\infty$, yet
the $m$ dependence is correctly captured by the continued result (\ref{I3}).
The same conclusion is drawn if other regularization schemes are employed,
such as a proper-time cutoff.  On the other hand, dimensional regularization
says $I_4$ is infinite, and indeed, there is no well-defined finite part of
that integral:
\be
I_4=\frac{2\pi^2}{(2\pi)^4}\int_0^\Lambda dl\,l^3\frac1{l^2+m^2}=
\frac1{16\pi^2}\left(\Lambda^2-m^2\ln\frac{\Lambda^2+m^2}{m^2}\right).
\ee

More generally, consider
\be
I(d,\alpha)=\int\frac{d^dl}{(2\pi)^d}\frac1{(l^2+m^2)^\alpha}.
\ee
For $\alpha>0$ and $d/2<\alpha$ we have
\be
I(d,\alpha)=\frac{m^{d-2\alpha}}{2^d\pi^{d/2}}\Gamma(\alpha-d/2).
\ee
This assigns the finite value to the integral with $\alpha=2$ and $d=5$:
\begin{subequations}
\be
I(5,2)=-\frac{m}{16\pi^2}\quad (\mbox{dimensional continuation}),
\ee
which is divergent when a momentum cutoff is used:
\be
I(5,2)=\frac{\Lambda}{12\pi^2}-\frac{m}{16\pi^2}, \quad \Lambda\to\infty.
\ee
\end{subequations}

The general conclusion is that dimensional continuation gives the correct
finite part of the Feynman graph. (The reader is invited to examine
other, more complicated examples.) Since that finite part
 is, at least in the cases
we consider in this paper, the only part that contains reference to
physical parameters, e.g., the radius of the sphere, we conclude that
it is effective in isolating the physically observable energy.

\acknowledgments I am grateful to the US Department of Energy for
partial support of this research.  I thank Bob Jaffe, 
Michael Bordag, Jack Ng, and Carl Bender for helpful conversations.


\end{document}